\documentclass{article}

\usepackage{fullpage}
\usepackage{graphics}
\usepackage{graphicx}
\usepackage{lscape}
\usepackage{amsmath}
\usepackage{natbib}
\usepackage{indentfirst}
\usepackage[utf8]{inputenc}
\usepackage[T1]{fontenc}
\usepackage{lineno}
\usepackage{wasysym} 
\usepackage{setspace}
\usepackage{authblk}

\newcommand{\bm}[1]{\mbox{\boldmath $ #1 $}}

\begin{document}

\linenumbers

\title{Bayesian Stable Isotope Mixing Models}
\author[1]{Andrew C. Parnell\thanks{Andrew.Parnell@ucd.ie}}
\author[2]{Donald L. Phillips\thanks{These authors contributed equally to this manuscript and are listed in random order}}
\author[3]{Stuart Bearhop$^\dagger$}
\author[4]{Brice X. Semmens$^\dagger$}
\author[5]{Eric J. Ward$^\dagger$}
\author[6]{Jonathan W. Moore$^\dagger$}
\author[7]{Andrew L. Jackson$^\dagger$}
\author[3]{Richard Inger}
\affil[1]{School of Mathematical Sciences (Statistics), Complex and Adaptive Systems Laboratory, University College Dublin, Ireland}
\affil[2]{U.S. Environmental Protection Agency, National Health \& Environmental Effects Research Laboratory, Oregon, USA}
\affil[3]{Centre for Ecology and Conservation, School of Biosciences, University of Exeter, UK}
\affil[4]{Scripps Institution of Oceanography, University of California, San Diego, 9500 Gilman Drive, La Jolla, California, USA}
\affil[5]{Northwest Fisheries Science Center, National Marine Fisheies Service, National Oceanic and Atmospheric Administration, Seattle, USA}
\affil[6]{Earth2Ocean Research Group, Simon Fraser University, Burnaby, Canada}
\affil[7]{School of Natural Sciences, Trinity College Dublin, Ireland}
\maketitle

\begin{abstract}
In this paper we review recent advances in Stable Isotope Mixing Models (SIMMs) and place them into an over-arching Bayesian statistical framework which allows for several useful extensions. SIMMs are used to quantify the proportional contributions of various sources to a mixture. The most widely used application is quantifying the diet of organisms based on the food sources they have been observed to consume. At the centre of the multivariate statistical model we propose is a compositional mixture of the food sources corrected for various metabolic factors. The compositional component of our model is based on the isometric log ratio (ilr) transform of \citet{Egozcue2003}. Through this transform we can apply a range of time series and non-parametric smoothing relationships. We illustrate our models with 3 case studies based on real animal dietary behaviour.
\end{abstract}

\section{Introduction}\label{intro}

Stable isotope analysis is an increasingly important tool in the study of ecological food webs. The technique utilises the fact that biologically active elements exist in more than one isotopic form. Generally the lighter isotopic form is much more abundant in the environment than the heavier form, although their relative abundance is altered by a range of biological, geochemical and anthropogenic processes. These processes produce isotopic gradients which are reflected in the tissues of plants and animals. Differences in relative abundance of these isotopes within a particular sample can be measured using a mass spectrometer and expressed as the ratio of heavy to light form, which can then be standardised against international reference samples and reported in the delta ($\delta$) notation as parts per thousand or per mil ($\permil$).\\

As a consumer's tissues are ultimately derived from the dietary sources they consume, it is possible to use stable isotope mixing models (SIMMs) to derive the assimilated diet of an individual, or a group of individuals, given the isotopic ratios of the consumers' tissues and  food sources \citep{Phillips2012}. A number of recent papers have proposed models to analyse such data, gathering over 1500 citations since their first introduction. More recently, the models proposed for such data are Bayesian. In this paper we review the different models proposed and bring them into an over-arching framework. We include three case studies ranging from the simple to the complex, together with JAGS \citep[Just Another Gibbs Sampler;][]{Plummer2003} code for their implementation\footnote{See \texttt{mathsci.ucd.ie/$\sim$parnell\_a/}}.\\

Generally the isotopic ratios of a sample of a consumer's tissues (e.g. blood, feathers, whiskers) are measured along with a representative sample of potential items from a consumer's diet. The consumer isotopic values are represented in the model as the convex combination of the source values where the coefficients in the simplex are the `dietary proportions'; strictly speaking they are the proportion of the consumers' dietary proteins obtained from the sources. Estimation of these dietary proportions is the main focus of our analysis. Most commonly the isotopic observations on the consumers and sources are multivariate with dimension 2. A thorough description of the uses of stable isotopes can be found in \cite{ingerbearhop2008}. \\ 

\begin{center}
\begin{figure}[!h]
\includegraphics[width=14cm]{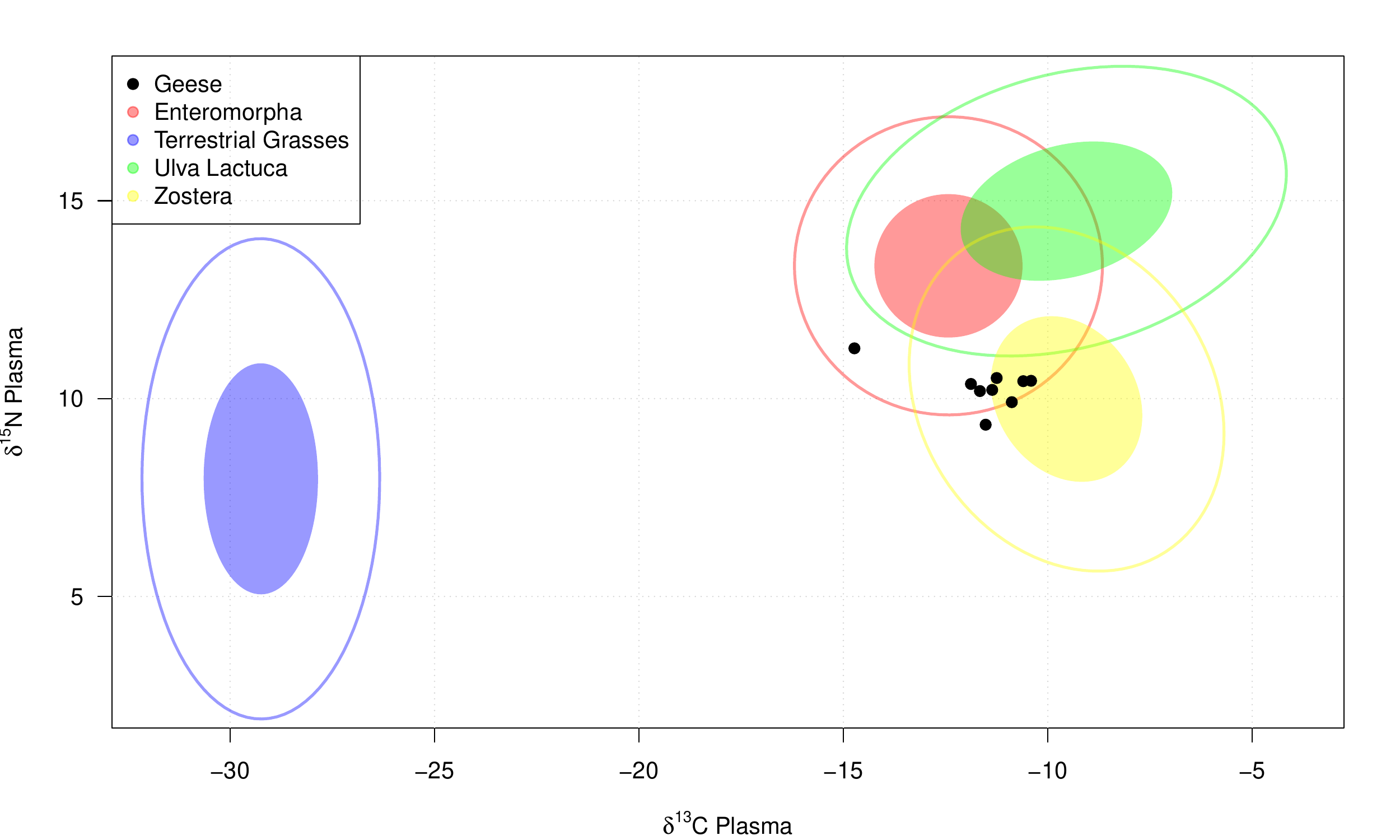}
\caption{Isospace plot of the geese data of case study 1 (see Section \ref{geese}). The consumer information is shown in the black filled circles, whilst the sources are shown as contours (90\% range) and filled ellipses (50\% range). The consumers seem to lie close to the \textit{Zostera} source so this is likely to form a substantial part of their diet. }
\label{GeeseIsoSpace}
\end{figure}
\end{center}

Once the isotopic data have been collected for both consumer and sources, it is usual to create an \textit{iso-space} plot which shows the consumer and source values. An example is shown in Figure \ref{GeeseIsoSpace}. It is desirable for the consumer values to lie within the fuzzy convex hull of the sources. However, a further phenomenon is often observed here, that of \textit{trophic enrichment}, 
whereby light isotopes are lost during the conversion of source proteins into consumer tissues. The isotopic values of the consumer (or equivalently the sources) are thus adjusted by a \textit{trophic enrichment factor} (TEF) which may vary by food source and consumer. These TEF corrections arise from laboratory studies, and thus contribute another set of (uncertain) data to our analysis. \\

The inference challenge involved in a SIMM is thus to estimate the dietary proportions whilst taking account of the uncertainty in the source and TEF values. Clearly not all of the consumers will eat exactly the same diet, so it is common to use a hierarchical model. Furthermore, covariates such as time or age may be available which are thought to influence the dietary proportions. The model we present in this paper takes account of these common features in a multivariate hierarchical Bayesian model.\\

The paper is organised as follows. In Section \ref{model} we introduce our general SIMM and show it may be extended to form part of a larger class of hierarchical compositional generalised linear models. In Section \ref{previous} we outline previous work in this area and show how each of these fits into our general SIMM. Section \ref{statissues} outlines the statistical issues concerning the formulation of such models and how they can be fitted. We outline three case studies in Section \ref{case}, showing how the model works in different situations depending on the information available. We discuss future directions in  Section \ref{discussion}.\\

\section{Model formulation}\label{model}

We first provide the notation we use to formulate our model. We suppose that there are $N$ consumers on $J$ isotopes, with $K$ sources. We outline the most important components of our model below:
\begin{itemize}
\item $Y_{ij}$ represents the isotope measurement on consumer $i$ for isotope $j$. We write $\bm{Y}_i$ as the $J$-vector of isotope values for consumer $i$
\item $s_{ijk}$ is the source value for consumer $i$ on isotope $j$ and source $k$. We write $\bm{s}_{ik}$ to be the $J$-vector of isotope source values for consumer $i$ on source $k$, and $\bm{s}_i$ to be the $J \times K$ matrix of source values for consumer $i$.
\item $c_{ijk}$ is the TEF value for consumer $i$ on isotope $j$ and source $k$. We write $\bm{c}_{ik}$ to be the $J$-vector of TEF values for consumer $i$ on source $k$, and $\bm{c}_{i}$ as the $J \times K$ matrix of TEF values related to consumer $i$.
\item $p_{ik}$ is the dietary contribution of source $k$ for consumer $i$. $\bm{p}_i$ is the $K$-vector of dietary proportions for consumer $i$. Estimation of these dietary proportions is the main focus of our analysis.
\item $\epsilon_{ijk}$ is a random noise term representing residual variation. We write $\bm{\epsilon_i}$ as the $J$-vector of residual terms for consumer $i$, and set $\bm{\epsilon}_i \sim N(0,\bm{\Sigma})$ with $\bm{\Sigma}$ a covariance matrix.
\end{itemize}

Using the notation above, we write our general model as:
\begin{eqnarray}
\label{mainmodel}
\bm{Y}_i \sim N(\bm{p}_i^T (\bm{s}_i+\bm{c}_i), \bm{\Sigma}).
\end{eqnarray}
We assume a hierarchical formulation so that $\bm{s}_{ik} \sim N(\bm{\mu}_k^s,\bm{\Sigma}_k^s)$ and $\bm{c}_{ik} \sim N(\bm{\mu}_k^c,\bm{\Sigma}_k^c)$ where the means and covariances of the sources and TEFs are estimated from the source and TEF experimental data.\\

Of particular interest is the modelling structure for the dietary proportions $\bm{p}$. We use an isometric log-ratio (ilr) approach as proposed by \cite{Egozcue2003}, though other transformations are available (see next two sections for further discussion). The transformation is written as:
\begin{eqnarray}
\label{eqnilr}
\bm{\phi}_i = \mathrm{ilr}(\bm{p_i}) = \bm{V}^T \log \left[ \frac{p_{i1}}{g(\bm{p}_i)}, \ldots,\frac{p_{iK}}{g(\bm{p}_i)} \right] \; \mathrm{with} \; g(\bm{p}_i) = \left( \prod_{i=1}^K p_{ik} \right)^{1/K}.
\end{eqnarray}
with $\bm{V}$ a $K-1 \times K$ matrix of orthonormal basis functions on the simplex. The inverse transformation $\bm{p}_i = \mathrm{ilr}^{-1} (\bm{\phi}_i)$ simply involves exponentiating and re-normalising the values. There are two consequences of working with the ilr. The first is that we now work in a $K-1$ dimensional space. The second is that there is no obvious link between the elements of $\phi_{ik}$ and $p_{ik}$, so we lose some degree of interpretability. We further parameterise the transformed proportions so that $\phi_{ik} \sim N(\gamma_{ik},\kappa_k)$ with $\kappa_k$ quantifying a random effect variance. Were we to use the centred log-ratio (clr) transform (equivalent to setting $\bm{V}=\bm{I}$; though see Section \ref{statissues} as to why we might not do this) then $\kappa_k$ would represent the consumer-level variance in diet for source $k$. Finally we set $\gamma_{ik}$ to be a mean term restricted in some fashion to allow for estimation. A multivariate prior for $\bm{\phi}_i$ would also be feasible.\\

In situations where covariates $\bm{x}_i$ are available, they are usually linked to the model through the dietary proportions. The covariates may take the form of age, sex, time or any other variables upon which diet is expected to depend. In certain cases (such as our third case study) both the diet and the sources are expected to be functions of a covariate. In the simpler case we apply the covariates by making $\gamma_{ik}$ functions of $\bm{x}_i$. In the more advanced case we additionally apply them to $\bm{\mu}_k^s$ and $\bm{\Sigma}_k^s$ .\\

We term the TEF data set $\mathcal{D}_c$ and the source data set $\mathcal{D}_s$, each consisting of collections of $J$-vectors of isotope ratios for each of the $K$ sources. A full Bayesian posterior distribution gives:
\begin{eqnarray}
\pi(\bm{p},\bm{\phi},\bm{\gamma},\bm{\kappa},\bm{\Sigma},\bm{\Sigma}_s,\bm{\Sigma_c},\bm{\mu}_s,\bm{\mu}_c,\bm{s},\bm{c}|\bm{Y},\bm{x},\mathcal{D}_c,\mathcal{D}_s) &\propto& \left[ \prod_{i=1}^N \pi(\bm{Y}_i|\bm{p}_i,\bm{s}_i,\bm{c}_i,\bm{\Sigma}) \right] \times \left[ \prod_{i=1}^N \pi(\bm{\phi}_i|\bm{\gamma}_i,\bm{x}_i,\bm{\kappa}) \right] \nonumber \\ 
& & \times \left[ \prod_{i=1}^N \prod_{k=1}^K \pi(\bm{s}_{ik}| \bm{\mu}_{k}^s,\bm{\Sigma}_k^s)\right] \times \left[ \prod_{i=1}^N \prod_{k=1}^K \pi(\bm{c}_{ik}| \bm{\mu}_{k}^c,\bm{\Sigma}_k^c)\right] \nonumber \\
& & \times \left[ \prod_{k=1}^K \pi(\bm{\mu}_{k}^s,\bm{\Sigma}_k^s|\mathcal{D}_s) \right] \times \left[ \prod_{k=1}^K \pi(\bm{\mu}_k^c,\bm{\Sigma}_k^c|\mathcal{D}_c) \right] \nonumber \\
& & \times \left[ \prod_{k=1}^K \pi(\kappa_K) \right] \times  \pi(\bm{\Sigma})
\end{eqnarray}

\begin{figure}[!h]
\begin{center}
\includegraphics[width=14cm]{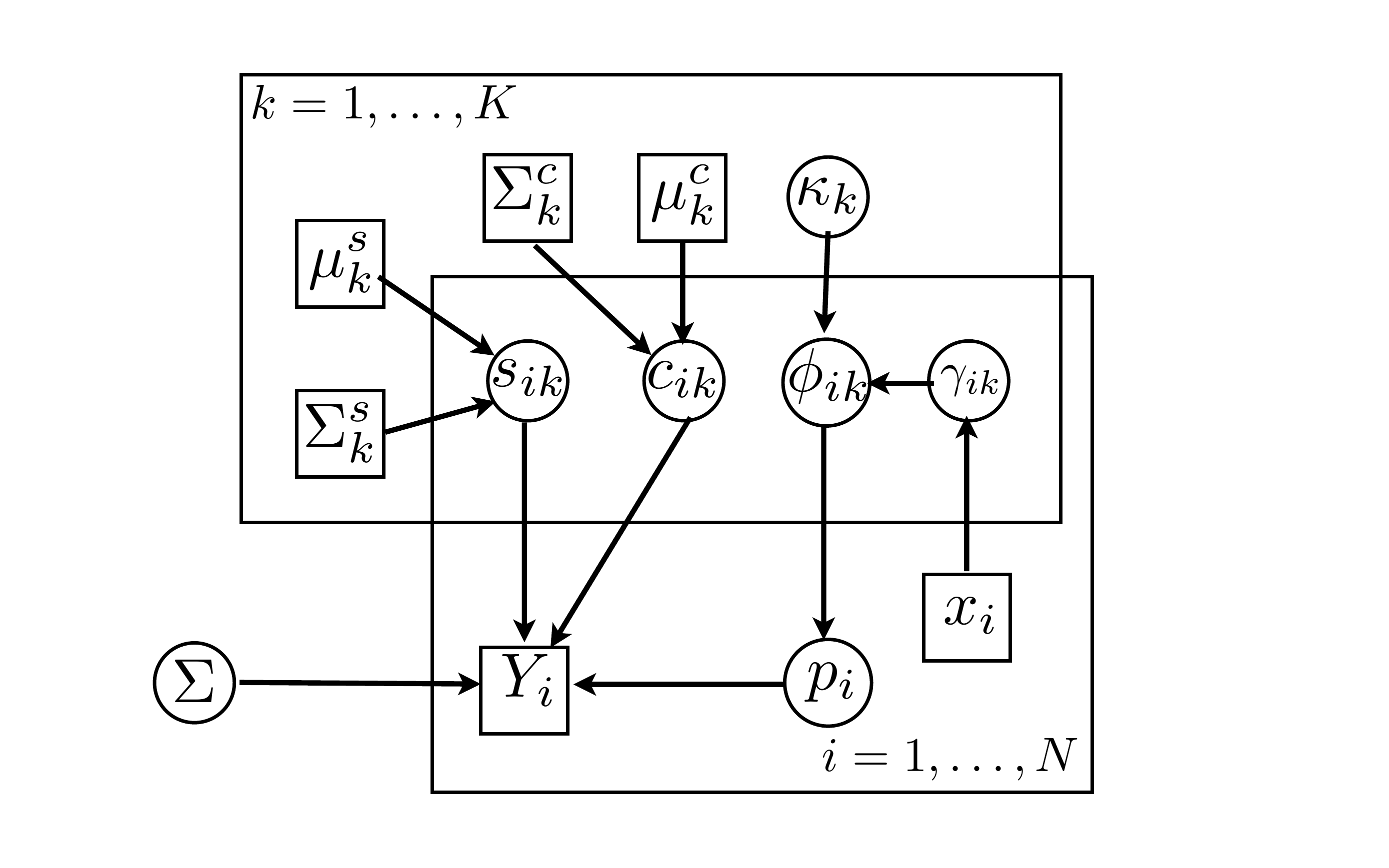}
\caption{A Directed Acyclic Graph (DAG) of our model. Circles indicate parameters to be estimated whilst squares indicate data. The arrows indicate the direction of information flow.}
\label{DAG}
\end{center}
\end{figure}

A Directed Acyclic Graph (DAG) is shown in Figure \ref{DAG}. The sub-models for the sources and TEFs, namely $\pi(\bm{\mu}_{k}^s,\bm{\Sigma}_k^s|\mathcal{D}_s)$ and $ \pi(\bm{\mu}_k^c,\bm{\Sigma}_k^c|\mathcal{D}_c) $ can be updated as part of the modelling steps, but we prefer a more efficient empirical Bayes approach \citep{Carlin2000} whereby we approximate $\bm{\mu}_k^s,\bm{\mu}_k^c,\bm{\Sigma}_k^s,\bm{\Sigma}_k^c$ by their sample estimates, thus cutting feedback and removing these from the updates and the posterior. Note that this has minimal effect on the source and TEF random effects $s_{ik}$ and $c_{ik}$ which are still updated as part of the modelling process. The prior distributions we give to $\bm{\Sigma}$ and $\kappa_k$ are vague Inverse-Wishart and Inverse-Gamma respectively. \\

\section{Previous work}\label{previous}

The earliest attempts at applying a mixing model framework to stable isotope data did not use probability as the basis for estimation. Initially, SIMMs were restricted to systems involving a single consumer (or the mean of multiple consumers), and where the number of isotopes and sources was arranged such that $J+1=K$. Such an arrangement yields a linear system with a single solution. \citet{phillipsgregg2001} provided propagation of error calculations for such a system in their IsoError model to establish confidence intervals around the estimates based on the variances of the consumer and source isotopic measurements. The method was expanded upon in IsoSource \citep{Phillips2003} to relax the $J+1=K$ restriction and allow for multiple sources but without explicit incorporation of source and consumer variability. Isosource works by simulating values of the dietary proportions $p$ on a grid to produce multiple valid solutions to the linear system. The valid solutions could be plotted in a histogram-like fashion, though they did not represent probability distributions, rather simply the range of values which might be plausible given the geometry of the system.\\

These initial attempts were formalised in a Bayesian fashion in the models MixSIR \citep{Moore2008} and SIAR \citep{parnelletal2008a}. The MixSIR model can be thought of as a simplification of Equation \ref{mainmodel} without explicit random effects across the dietary proportions, and with $\bm{\Sigma}$ set to zero. The sources and TEFs are treated as independent across isotopes and are given fixed values for their mean and variance. The dietary proportions are given independent Beta-distributed priors or, in a later version, a Dirichlet distribution. Since the dietary proportions are the only parameters in the model, it can be fitted extremely efficiently using Importance Resampling \citep[e.g.][]{Robert2005} on a grid encompassing the range of proportion values. Updated versions of the MixSIR model have included random effects in the dietary proportions through the clr transform, and also have allowed for hierarchical models to be fitted, most elegantly in capturing familial relationships affecting the diet of gray wolves in British Columbia \citep{Semmens2009}. These latter advanced versions of MixSIR are fitted using the JAGS software \citep{Plummer2003}. \\

The SIAR model \citep{Parnell2010} is in many ways similar to the basic MixSIR model (and thus still a simplification of Equation \ref{mainmodel}) though includes a residual component which is treated as independent between isotopes and given an Inverse Gamma prior. The model also allows for concentration dependencies ($q_{kj}$, vectorised as $\bm{q}$) which quantify the proportion of the isotope in the given food source \citep{Phillips2002}. They can be added in to our model by replacing $\bm{p}_i$ in Equation \ref{mainmodel} with $\mathcal{C} (\bm{p}_i \oplus \bm{q})$ where $\mathcal{C}$ is the simplex closure operator and $\oplus$ is the simplex perturbation \citep{Egozcue2003}. Usually $\bm{q}$ are either fixed or given a suitably informative prior distribution. The SIAR model is fitted using standard MCMC with Metropolis-Hastings steps to update the dietary proportions.\\

More recently, the IsotopeR model has been introduced which extends the SIAR/MixSIR models to a multivariate setting (both sources and TEFs are multivariate normal) and partitions the residual covariance $\bm{\Sigma}$ into a mass spectrometer calibration error and that of residual error. They further allow for the sources, TEFs and dietary proportions to be random effects with the latter obtained through the clr transform (see next section for further discussion). The model is fitted in JAGS \citep{Plummer2003} but does not allow for covariate information or for the estimation of the dietary random effect variance.\\

Aside from explicit SIMM model development, recent focus has also been on the performance of SIMMs in non-ideal conditions, most notably with respect to the characterisation of source and TEF values by \citet{Bond2011}. Clearly it is absolutely vital that food sources are not excluded from the model as they will yield biased dietary proportions. Similarly, the estimation of the TEF values must be conducted appropriately or there will be some extra uncertainty in the estimated dietary proportions. In related work, \citet{Ward2011} considered the problem of (dis)aggregating sources and its effect on the resulting estimates. For example, if a consumer only eats part of a food source it may be hard to obtain isotopic values from just that part. Similarly if two sources, though different species, lie in the same location in iso-space it may be impossible for the model to determine the difference in their dietary consumptions. Thus on occasion it may be pertinent to aggregate sources without any loss of information. Alternatively the aggregation can be accomplished with fewer assumptions if it is calculated \textit{a posteriori} by combining the negatively correlated dietary proportions.\\

Lastly, it should be noted that there are strong connections between SIMMs and the `end member' analysis used by geologists to determine the composition of e.g. river sediment. In such cases the sources are usually known with minimal error and the challenge is to estimate the proportional contributions of different sediment sources. A number of different methods have been proposed, e.g. \citet{Soulsby2003,Brewer2011} (and references therein). Ours most closely resembles that of \cite{Palmer2008}, though they use the alr transformed proportions (see below for definition) which are given a spatial prior distribution.\\

\section{Statistical issues in SIMMs}\label{statissues}

The models we fit use ideas from Bayesian hierarchical modelling \cite[e.g.][]{Gelman2003} and compositional data anaylsis \citep{Aitchison1986}. BHM is now part of the standard toolbox of Bayesian statistics and we do not discuss them further here. Of more interest however is the compositional structure applied to the dietary proportions as this can strongly affect the behaviour of the posterior distribution. A recent review of the state of compositional data analysis can be found in \citet{Pawlowsky-Glahn2011}. Our models differ fundamentally from many standard problems as our compositions are not observed directly but are latent parameters to be estimated, constrained by their geometrical position in iso-space and the covariates upon which they may depend.\\

The starting point for the early Bayesian SIMMs models was that of the Dirichlet distribution, being perhaps the simplest valid distribution on the simplex. The usual prior distributions used were either flat where all Dirichlet parameters are set to 1 or the Jeffreys prior where all are set to $1/K$. Unfortunately, as is well known \citep[e.g.][]{Aitchison1986}, the Dirichlet distribution suffers from a very rigid sub-compositional independence assumption. This is not necessarily a problem when used as a prior distribution with fixed components as the posterior distribution may well show interesting sub-compositional properties. However, if dependence is to be modelled through hyper-parameters of the Dirichlet distribution (e.g. with covariates) this restriction will remain in the posterior.\\

A number of extensions to the Dirichlet have been proposed \citep[e.g.][]{Wong1998}, but we focus here on the logistic-normal transformations of \citet{Aitchison1986} and \cite{Egozcue2003}, through which more flexible sub-compositional dependence can be obtained. The simplest of these is perhaps the additive log-ratio (alr), where $g(\bm{p}_i)$ in Equation \ref{eqnilr} is set to be one of the chosen proportions (which is then removed from the composition) and $\bm{V}=\bm{I}$. However, this can perform poorly when there is no obvious choice of denominator and is not permutation-invariant. The centred log-ratio (clr; defined in Equation \ref{eqnilr} when \bm{V}=\bm{I}) removes the need for a choice of denominator in the log ratio but produces a covariate matrix of rank $K-1$. It is also not sub-compositionally coherent \citep[see][for further discussion of these terms]{Pawlowsky-Glahn2011}. Finally the isometric log-ratio (ilr) uses orthonormal basis functions in the simplex to obtain coordinates that are isometric and satisfy the usual compositional requirements of coherence and permutation/subcomposition invariance. The choice of basis functions is somewhat subjective; we follow the method in \cite{Egozcue2003}. From a Bayesian perspective, the latent compositional parameters are more easily identifiable when working with the ilr.\\

Once a suitable transform has been chosen, it is feasible to include covariates or perform any of the traditional multivariate analysis techniques. Numerous examples can be found, including a geo-statistical framework \citep{Tolosana-Delgado2011}, discrete time series \citep{Barcelo-Vidal2011}, or spectral methods \citep{Pardo-Iguzquiza2011}. A popular topic is that of zero compositions or zero-inflation \citep[e.g.][]{Butler2008} which is not a severe issue in SIMMs because we know from experimental observation that all dietary sources are consumed.\\

\section{Case studies}\label{case}
 
We now present three case studies and show how our general model can be used in each of the scenarios. In the first case study we analyse the diet of a small sample of geese from data previously studied by \cite{ingeretal2006b}. This first case study uses the model as proposed in Section \ref{model}, and can be seen as a small extension to that of \cite{Hopkins2012}. The second case study extends the geese model to allow for the inclusion of covariates or basis function models. Our final case study includes a compositional time series component where the sources and consumers are observed at different time points. In this final case study the data are swallows consuming chironomid midges, other freshwater invertebrates and terrestrial invertebrates. In all cases we run the models using the JAGS software and check convergence using the  coda package \citep{Plummer2006} and the Brooks-Gelman-Rubin diagnostic \citep{gelmanrubin1,brooksgelman1}.\\

\subsection{Case study 1}\label{geese} 

Our first case study contains 9 ($\delta^{13}$C, $\delta^{15}$N) pairs of isotopic values taken from the blood plasma of Brent geese sampled on 26th October 2003. The food sources are \textit{Zostera} spp, terrestrial grasses, \textit{Ulva lactuca} and \textit{Enteromorpha} spp. Our empirical Bayes estimates for the sources are given in Table \ref{GeeseSources}.  The TEFs were taken from values in the literature \cite[see references in][for more details]{ingeretal2006b} so that $\bm{\mu}_k^c = (1.63,3.54)^T$ and $\bm{\Sigma}_k^c=\left[ \begin{array}{cc} 1 & 0 \\ 0 & 1 \end{array} \right]$ for all $k$. For a simple random effects formulation we set $\gamma_{ik}=0$ over all $i$ and $k$. An iso-space plot for the data (where the sources have been corrected by the TEFs) is shown in Figure \ref{GeeseIsoSpace}. The JAGS model was run for 3 chains over 50,000 iterations, removing 10,000 for burn-in and thinning by a factor of 20.\\

\begin{table}[!h]
\begin{center}
\begin{tabular}{lcccc}
\hline
Source & \textit{Enteromorpha} & Terr Grasses & \textit{Ulva lactuca} & \textit{Zostera} \\
\hline
$\bm{\mu}_s^k$ & $(-14.06,9.82)^T$ & $(-30.88,4.43)^T$ & $(-11.17,11.2)^T$ & $(-11.17,6.45)^T$ \\
$\bm{\Sigma}_s^k$ & $\left[ \begin{array}{cc} 1.37 & 0 \\ 0 & 1.37 \end{array} \right]$ & $\left[ \begin{array}{cc} 0.41 & 0 \\ 0 & 5.15 \end{array} \right]$ & $\left[ \begin{array}{cc} 3.83 & 0.85 \\ 0.85 & 1.24 \end{array} \right]$ & $\left[ \begin{array}{cc} 1.48 & -0.56 \\ -0.56 & 2.16 \end{array} \right]$ \\
\hline
\end{tabular}
\caption{Estimates of the source means and covariance matrices for the geese data of case study 1.}
\label{GeeseSources}
\end{center}
\end{table}

\begin{center}
\begin{figure}[!h]
\begin{tabular}{cc}
\includegraphics[width=7.5cm]{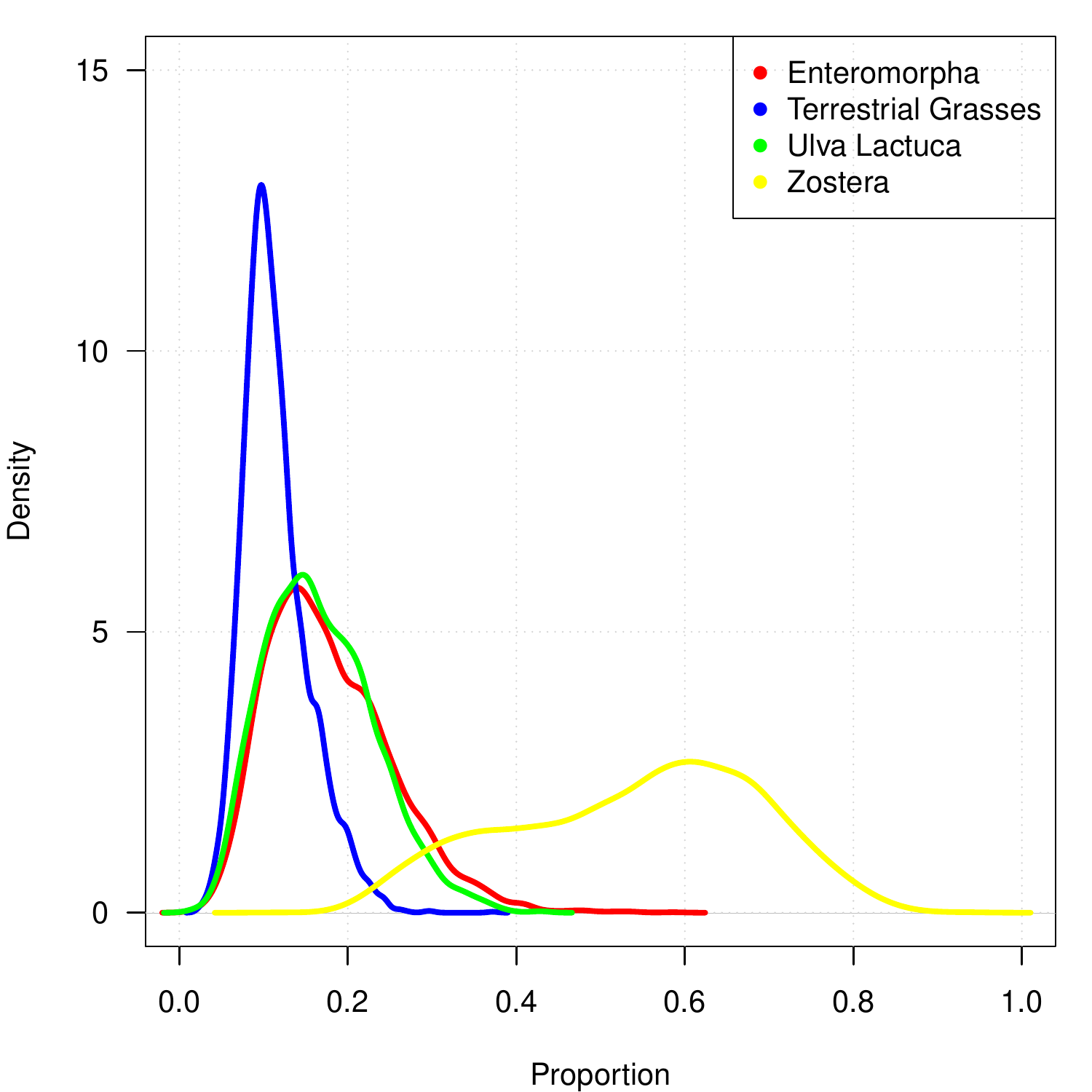} & \includegraphics[width=7.5cm]{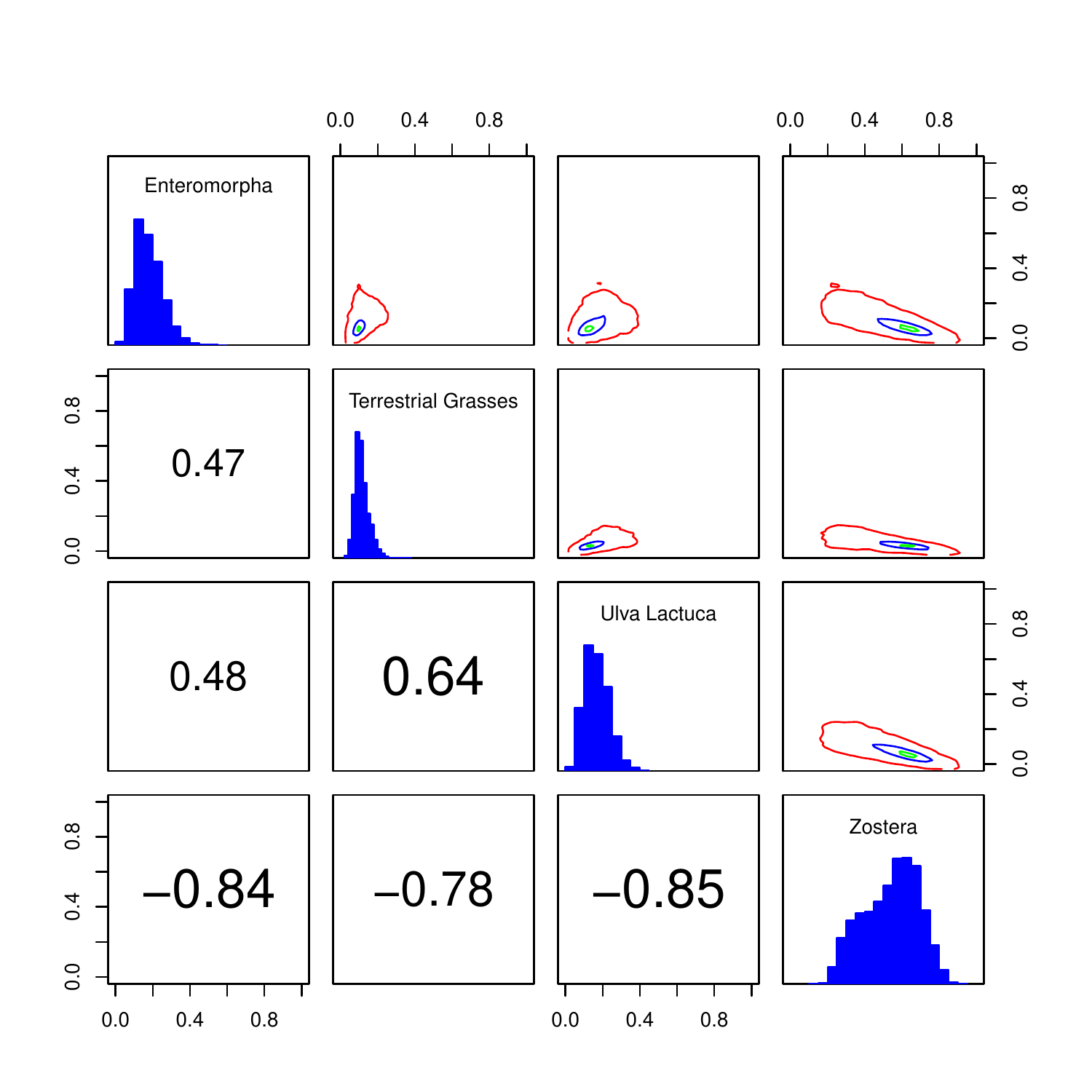}
\end{tabular}
\caption{Left panel: Density plot of mean dietary proportions for the geese data of case study 1. Right panel: matrix plot of the posterior dietary proportions obtained from the geese data. The upper-diagonal shows a contour plot, the diagonal a histogram, and the below-diagonal the correlation between the different sources.}
\label{GeeseOutput}
\end{figure}
\end{center}

A density plot of the mean dietary proportions is shown in the left panel of Figure \ref{GeeseOutput}. They can be seen to compare favourably with the simpler SIAR model \citep[see the function \texttt{siardemo} in][]{parnelletal2008a}, though in this case we have extra information covering individual dietary estimates, as well as improved estimation of the source and TEF random effects. In particular, the flexibility of the hierarchical formulation through the ilr transform allows for some multi-modality in the posterior distributions. The right panel of Figure \ref{GeeseOutput} shows a matrix plot of the joint behaviour of the dietary proportions. These can be useful in determining unavoidable model inadequacy, for example when it is impossible to ascertain which food sources are being consumed together. A strong negative correlation indicates that the food sources are indistinguishable. For example the strong negative correlation between \textit{Zostera} and \textit{Ulva lactuca} indicates that, whilst it is clear they are consuming mainly \textit{Zostera}, the balance between the two cannot be exactly determined. \\

There are many other useful statistics we can calculate here quite simply from the posterior distributions. In particular, we can focus on individual level variation by calculating, for example, the probability that an individual consumes more \textit{Zostera} than another. Similarly with access to the variance parameters $\kappa_k$ we can determine whether there is more variation amongst consumers in some sources rather than others. Lastly, it is often desirable to perform model comparison diagnostics to determine whether certain parts of the model can be removed without a detrimental effect on prediction. However, we do not perform any further analysis on this data set, preferring instead to use these tools when analysing the more sophisticated data below.\\

\subsection{Case study 2}\label{geesecovar}

\begin{figure}[!h]
\begin{center}
\includegraphics[width=14cm]{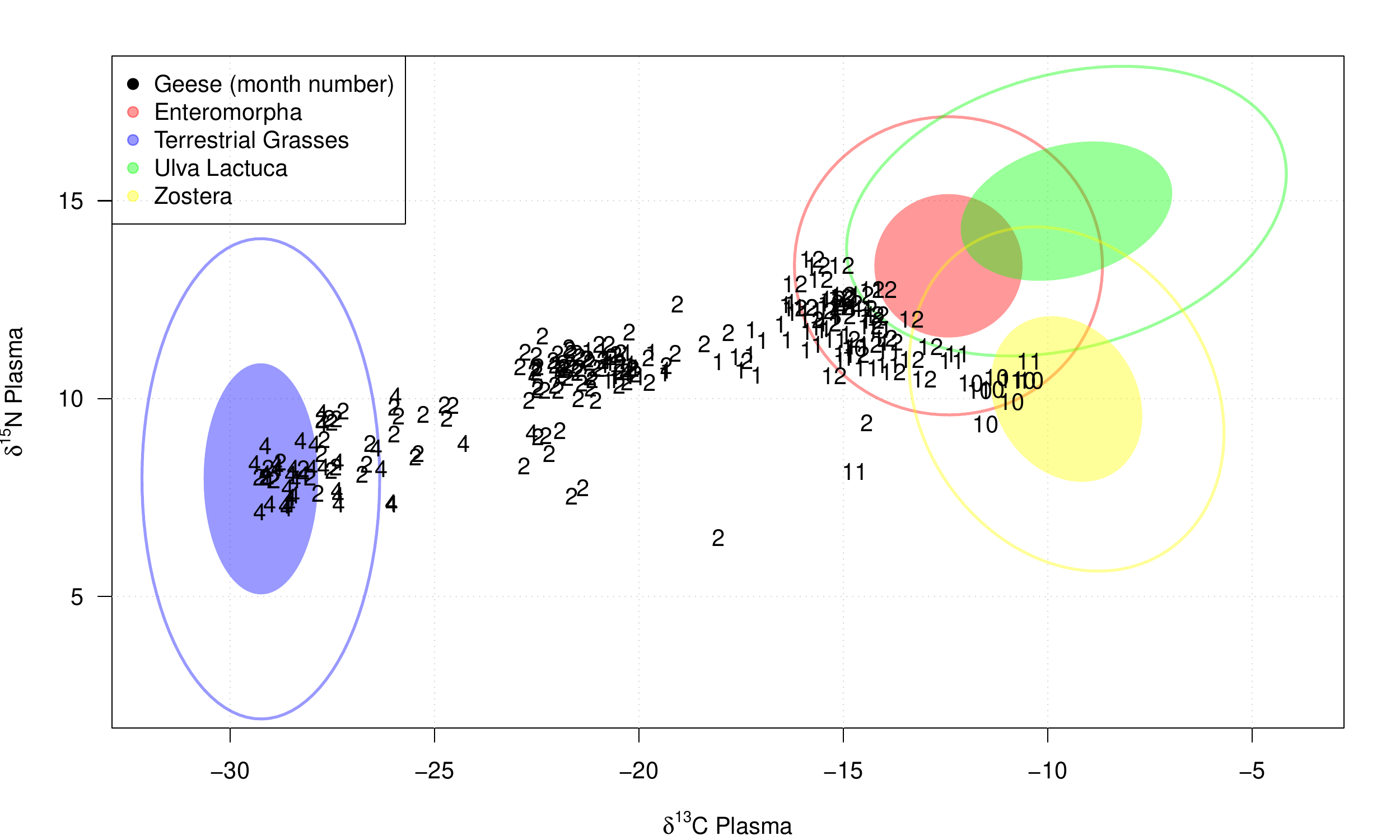}
\end{center}
\caption{Isospace plot of the full goose data set where the consumers are labelled by month. The data in case study 1 correspond to the data shown for month 10. Note that the sources and the TEFs are unchanged from Figure \ref{GeeseIsoSpace}.}
\label{Geese2IsoSpace}
\end{figure}

We now extend our Goose model to a larger data set of 248 observations collected over the period of October 2003 to April 2005 of which the previous case study was just a small part. The sources and TEFs are believed to be stable over the course of the study so the source means and covariances are the same as Table \ref{GeeseSources}. The diet of the geese will however vary during the season due to variations in abundance of food sources along with social and demographic factors. An iso-space plot for the full data is shown in Figure \ref{Geese2IsoSpace}. The iso-space plot appears to show that the diet during October to be focussed mainly on \textit{Zostera}, moving on to \textit{Ulva lactuca} and/or \textit{Enteromorpha} during November/December. In January and February the diet appears to be relatively mixed, but focussing almost solely on Terrestrial grasses around April. In addition to the stable isotope information we have covariates which state the Goose's sex and whether they are juvenile or adult.\\

We consider 6 possible models for the dietary behaviour, accounting for the covariates. In each case we set $\bm{\gamma}_{ik} = \bm{X}_{i}^T \bm{\beta}_k$ where $\bm{X}_{i}$ is an $L$-vector of covariates or basis functions for consumer $i$, and $\bm{\beta}_k$ is an $L$-vector of regression parameters associated with $\phi_k$. Note that, when using the ilr transform, the parameters $\bm{\beta}_k$ do not have any association with source $k$. With the extra parameters in the model, convergence of the MCMC algorithm is a problem because the parameters are often highly correlated across sources, though this is somewhat lessened with appropriate choice of $\bm{V}$ in the ilr. We find it helpful to re-parameterise with Helmert contrasts across sources so that $\bm{\gamma}_{i1} = \bm{X}_{i}^T \bm{\beta}_1$ and $\bm{\gamma}_{ik} = \bm{X}_{i}^T (\bm{\beta}_1+\bm{\beta}_k)$ for $k\ge 2$. We use the first source (\textit{Enteromorpha}) as $k=1$ as this seems to be consumed throughout the season though this is obviously not known in advance so does involve some degree of trial and error.\\

In all cases the parameters $\beta$ are given vaguely informative $N(0,10)$ distributions as a value of $|\phi|$ in excess of 10 is likely to yield dietary proportions near 100\% (though again this depends on the correlation structure in the data set). We compare the different models using the Deviance Information Criterion \citep[DIC;][]{spiegelhalteretal2002} and, for the final chosen model, posterior predictive distributions of the data. We use two versions of the DIC, the standard $p_D$ method of \citet{spiegelhalteretal2002}, and the $p_V$ method of \citet{Plummer2008} which estimates the optimism as half the variance of the deviance, and thus penalises complex models more harshly. Due to the extra complexity in the models, we run them with 3 chains for 200,000 iterations, removing 20,000 for burn-in and thinning by 90.\\

The first model we try involves no covariates and is thus the same as that in Section \ref{geese}. The second model includes an intercept term and time as a simple linear covariate. The third model replaces linear time with  a single harmonic component so that $X_{i}^T = \left[ 1,\cos \left( \frac{2\pi t_i}{365} \right), \sin \left( \frac{2\pi t_i}{365} \right)\right]$ where $t_i$ is the Julian day. The fourth, fifth and sixth models are expansions of model 3 to include juvenile/adulthood (model 4), sex (model 5) and also their interaction (model 6) as covariates. Table \ref{tabledic} shows the different models and the associated DIC values. Both versions of DIC seem to prefer models with the harmonic covariate, and also the addition of either sex or adulthood.\\

\begin{figure}[!h]
\begin{center}
\begin{tabular}{cc}
\includegraphics[width=8cm]{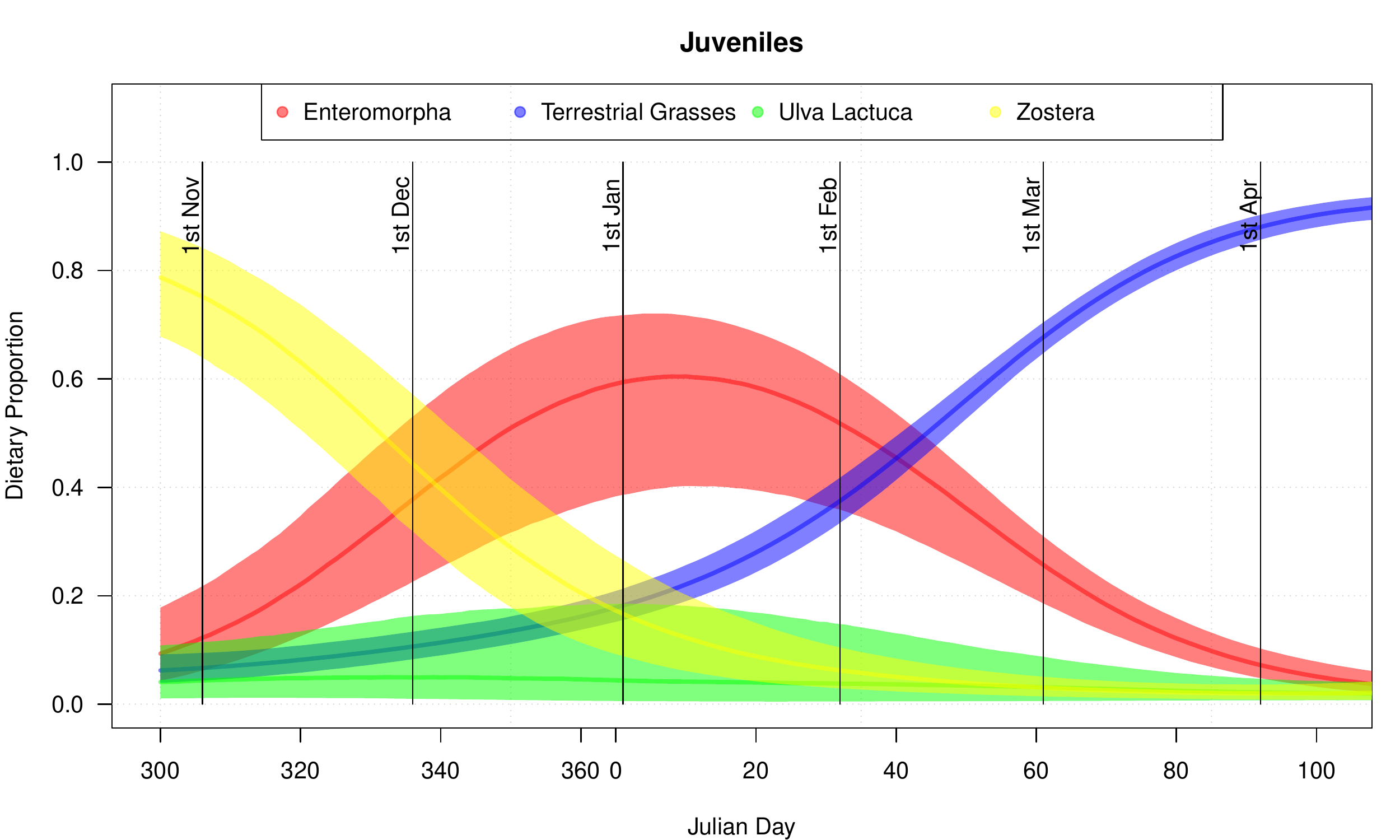} & \includegraphics[width=8cm]{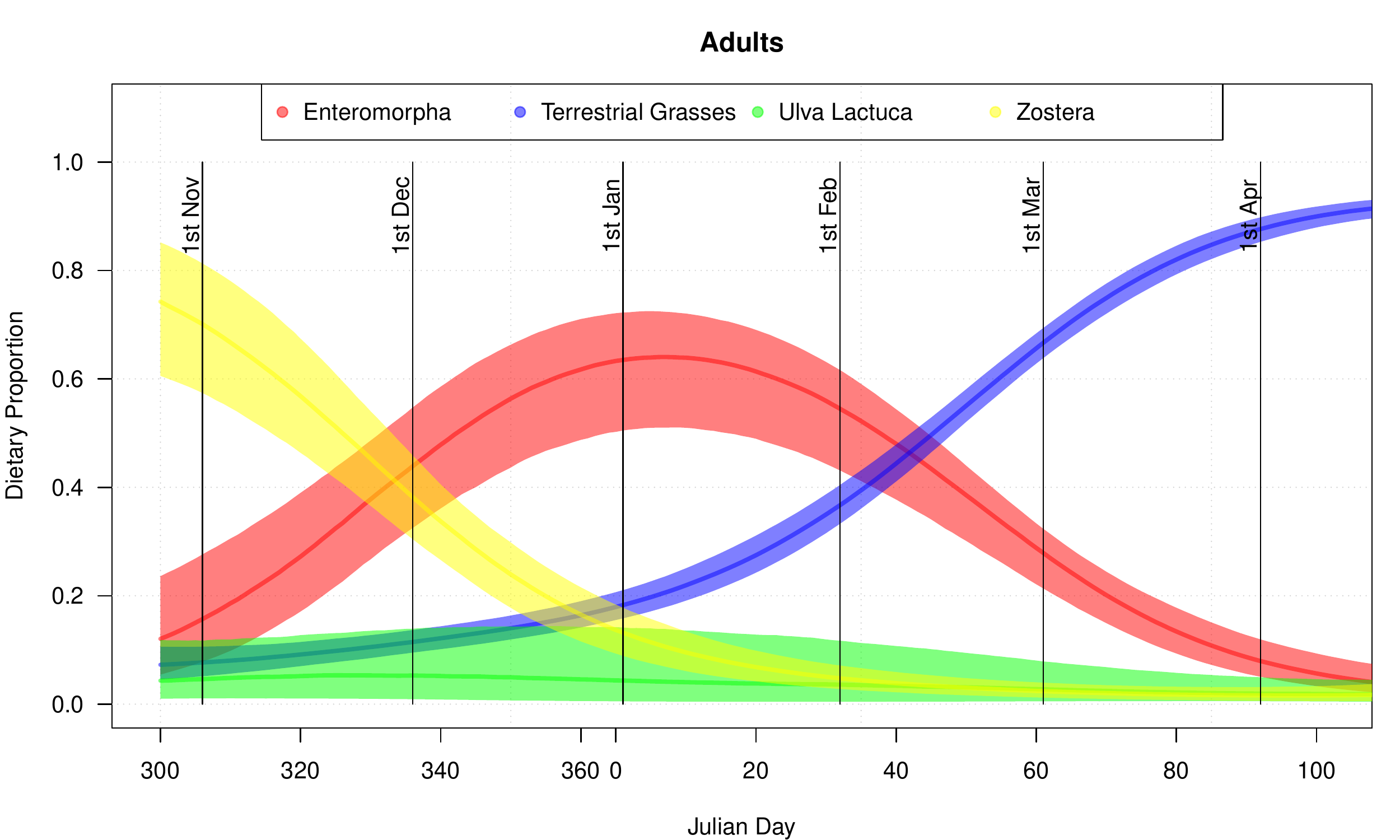}
\end{tabular}
\end{center}
\caption{Plot of proportion values against Julian day for model 3. The left panel shows estimates for juveniles whilst the right panel shows adults. The solid lines show median estimates, the outer of the polygons show 90\% credibility intervals. The Geese appear to focus on \textit{Zostera} around November before moving on to \textit{Enteromorpha} and then terrestrial grasses. There is slightly more uncertainty in the juveniles than the adults.}
\label{Geese2Harmonic}
\end{figure}

\begin{figure}[!h]
\begin{center}
\includegraphics[width=14cm]{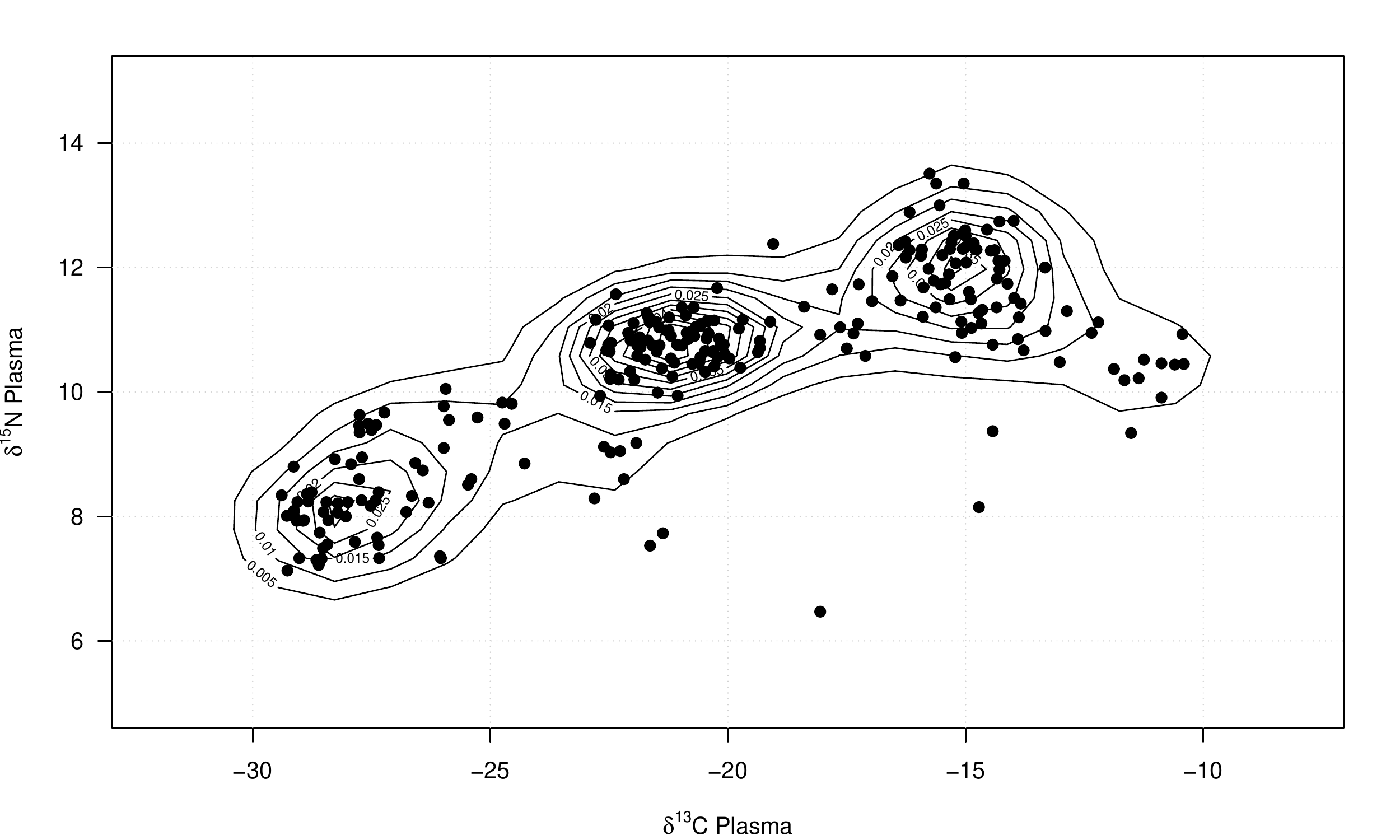}
\end{center}
\caption{Plot of the predictive distribution of the data from chosen model. The observations are shown as filled circles, whist the posterior predictive density is represented by the contours.}
\label{Geese2Predictive}
\end{figure}

Figure \ref{Geese2Harmonic} shows the relationship between Julian day and dietary proportion for the different sources for model 4. The switch from  \textit{Zostera}, to \textit{Enteromorpha}, to terrestrial grasses is very clearly seen in both juveniles and adults, though the uncertainty in juveniles is slightly higher, especially with respect to \textit{Enteromorpha}. Figure \ref{Geese2Predictive} shows the posterior predictive distribution of the data under model 4. The model seems to predict the data well, the three modes corresponding to the main sampling times of October, February and April. \\

\begin{table}[!h]
\begin{center}
\begin{tabular}{llrr}
\hline
\hline
Model & Covariate(s) & DIC (using $p_V$) & DIC (using $p_D$) \\
\hline
1 & None & 26907 & 1210.0 \\
2 & Julian day (linear) & 16957 & 524.3 \\
3 & Julian day (harmonic) & 16683 & 385.1 \\
\textbf{4} & \textbf{Julian day (harmonic), Juvenile/Adult} & \textbf{7551} & 393.6 \\
\textbf{5} & \textbf{Julian day (harmonic), Sex} & 8600 & \textbf{382.8} \\
6 & Julian day (harmonic), Juvenile/Adult, Sex, Interaction & 10812 & 382.9 \\
\hline
\hline
\end{tabular}
\end{center}
\caption{Table of models and model selection criteria. The models with the lowest DIC are shown in bold.}
\label{tabledic}
\end{table}

\subsection{Case study 3}\label{swallows}
 
Our last case study concerns the dietary behaviour of barn swallows (unpublished data). The data are stable isotope ratios of blood plasma samples from birds captured between May and August in 2009. Again, we use Julian day to determine their behaviour over time. However, in this scenario the sources (chironomid midges, other freshwater invertebrates, and terrestrial invertebrates) are also expected to change over time, and may have been observed on different days to that of the swallows. We thus expand our model to include a temporal component on the source vector as well as that of the dietary proportions. In each case, we use P-splines \citep{Eilers1996} to explain the suitably flexible behaviour. Other time series methods, such as random walks or Levy processes, may also be appropriate. We write continuous time as $t_i$ so that the consumers are now $\bm{Y}(t_i)$. We set $\gamma_{k}(t_i) = \bm{X}_{i}^T \bm{\beta}_k$ where now $\bm{X}_i$ is a an $L$-vector of cubic B-spline basis functions evaluated at time point $t_i$ and $\beta_k$ are weights for each basis function on source $k$. The P-spline formulation is completed by giving a random walk prior such that $\beta_{lk}-\beta_{l-1,k} \sim N(0,\tau_k^{-1})$ where $\tau_k$ is a roughness parameter associated with source $k$ and given a weakly informative $Ga(2,1)$ prior. \\

\begin{figure}[!h]
\begin{center}
\includegraphics[width=14cm]{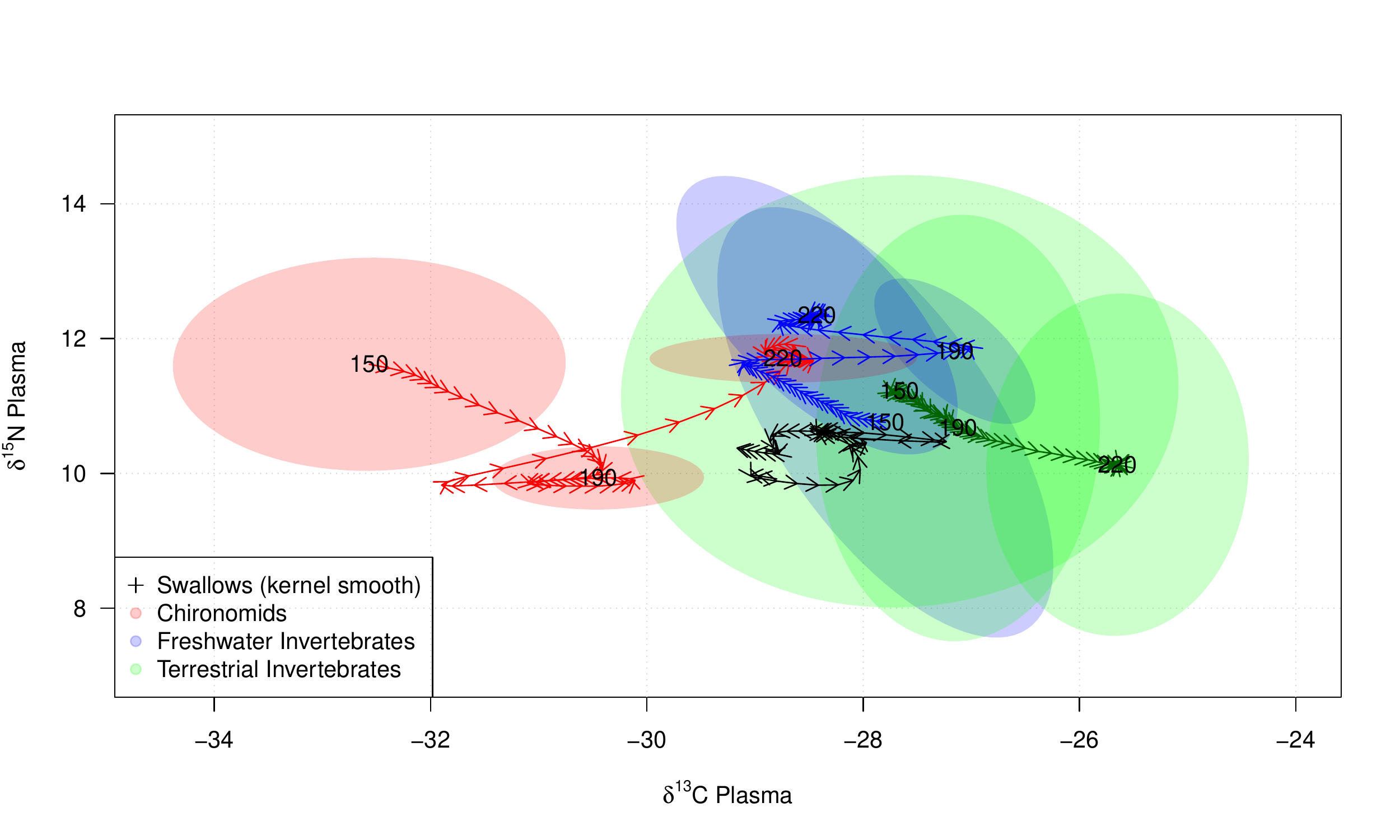}
\end{center}
\caption{Isospace plot of the swallows data. The source spline model has been run to obtain estimates of the source means and covariances throughout the study period. Arrows indicate the direction of movement over time for the sources and the swallows. The Julian day and the 50\% standard ellipses are given for Julian days 150, 190 and 220. Note that the chironomids' $\delta^{13}$C values increase over time from the start of the study period. A similar occurrence can be seen in the terrestrial invertebrates. The data are shown as kernel-smoothed estimates of the swallows' isotope data again over Julian day, starting at 150 and ending on day 220. The swallows can be seen to also increase their $\delta^{13}$C values up until around day 180 whereupon the $\delta^{13}$C returns towards its original value. This plot should be read in conjunction with Figure \ref{SwallowsSpline}.}
\label{SwallowsIsoSpace}
\end{figure}

The sources are now described by a multivariate spline model so that source data pairs, denoted $s'_{jk}(t)$ for the source experimental data at time $t$ on isotope $j$ and source $k$, are given $N \left( \left[ \bm{X}^T \bm{\beta}_1',\ldots, \bm{X}^T \bm{\beta}_J' \right]^T, \bm{\Sigma}'(t) \right)$ independently for each source $k$. The number of observations for each source is likely to be different, and certainly not equal to the number of consumer observations $N$. Here, the spline parameters $\bm{\beta}_j'$ determine the mean behaviour of the sources over time on isotope $j$. The $J \times J$ variance matrix $\bm{\Sigma}'(t)$ is also allowed to change over time with diagonal elements given log-splines: $\log(\Sigma_{jj}' (t)) \sim N(\bm{X}^T\bm{\beta}_{\Sigma},\kappa_\Sigma)$. The cross isotope covariance $\Sigma_{12}'$ is parameterised through a single correlation parameter for each source, denoted $\rho_k$, and does not change over time. A spline could also be used here (for example on the arctangent of $\rho_k$) but this was not found to improve the fit. The source spline model was run for each of the three sources in turn and used to calculate Maximum a Posteriori (MAP) estimates of $\bm{\mu}_k^s(t)$ and $\bm{\Sigma}_k^s(t)$ for all times $t$ at which consumer data were available. An iso-space plot of the swallows data is shown in Figure \ref{SwallowsIsoSpace}.\\

We use the source model predictions in our standard SIMM with the spline formulation on the proportions as outlined above. An alternative model where all parameters are estimated simultaneously would be possible, if rather slow. Instead, we retain the cutting feedback assumption so that the source and dietary proportions are run separately. For both the source and the dietary proportion models we run for 200,000 iterations, removing 20,000 for burn-in and thinning by 90. For both models, we use 25 knots, which seems to cover the flexibility in the data adequately.\\

\begin{figure}[!h]
\begin{center}
\includegraphics[width=14cm]{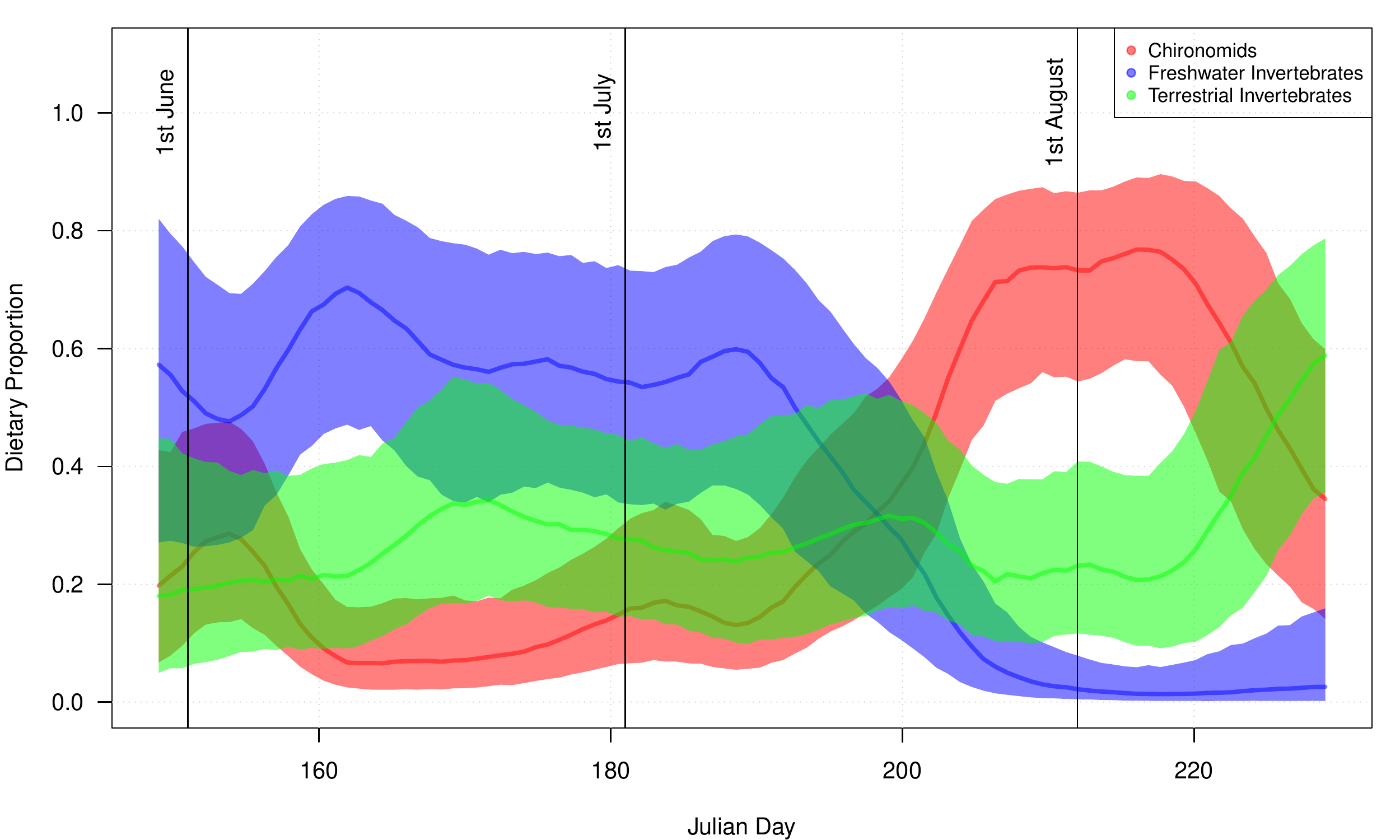}
\end{center}
\caption{Plot of dietary proportion values against Julian day for the swallows data of Case Study 3. The solid lines show median estimates, whilst the filled polygons show 90\% credibility intervals for each source.}
\label{SwallowsSpline}
\end{figure}

\begin{figure}[!h]
\begin{center}
\includegraphics[width=14cm]{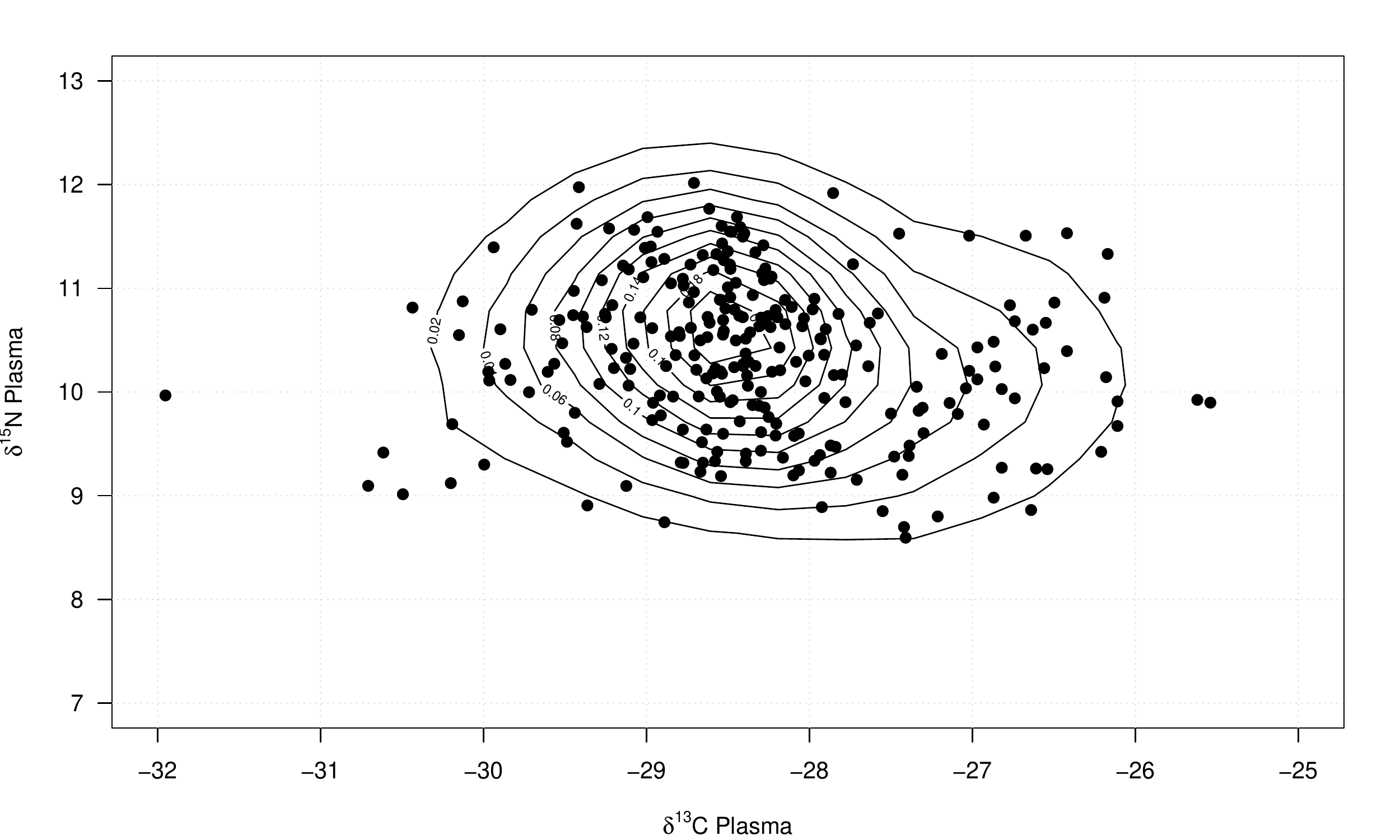}
\end{center}
\caption{Predictive distribution of the data for the swallows example. The observations are shown as filled circles, whist the posterior predictive density is represented by the contours.}
\label{SwallowsPredictive}
\end{figure}

Figure \ref{SwallowsSpline} shows the posterior dietary proportion estimates over Julian day for the swallows, predicted from the resulting spline parameter estimates. The results clearly indicate that they are feeding on mainly fresh water invertebrates during the early part of the data before concentrating on chironomids around the start of August. Figure \ref{SwallowsPredictive} shows the predictive distribution of the data under this model. The fit appears satisfactory. \\

\section{Discussion and future directions}\label{discussion}

The SIMM formulation outlined in Section \ref{model} allows for a rich framework upon which to include a variety of other statistical structures. The basis of such models is a mixture compositional structure applied to the dietary sources. In the case studies above we have illustrated how some simple regression and smoothing models might be included. The results seem useful and allow for many interesting findings which would not have otherwise been possible without, e.g. the time series or spline components.\\

The main challenges in such models are that the source and TEF values are fully and correctly characterised. It is a simple geometrical exercise to verify that if sources are missing from the data set, or that the TEF means and errors are poorly estimated, the dietary proportion estimates will be biased. This bias is compounded by the compositional nature of the problem where, if one dietary proportion is poorly estimated, then the others may well be too. There is no statistical test for missing sources; we rely on the ability of the ecologist to observe the system adequately. This may be particularly important if multiple organisms and sources are to be analysed simultaneously in a dietary network. A single missing source may cause biases across the dietary proportion estimates. Such models are not to be encouraged unless there is very strong evidence that no sources are missing and that TEFs are estimated correctly. \\

It is reasonably straightforward to expand our approach for other values of $J$ where differing numbers of isotopes may be available for analysis. It should be noted, however, that if $J\ge3$ it becomes harder to determine model fit, especially as no obvious iso-space plot can be created. The solutions to this problem may lie in closer scrutiny of predictive distributions, and some subjective judgement over the size of the residual covariance matrix $\Sigma$. In such scenarios extra care is required in reporting the output of the SIMM.\\

There are further opportunities for expansion of the models. Some of these may include:
\begin{itemize}
\item The simultaneous inclusion of multiple tissue varieties. It is often the case that different tissues are sampled on the same consumer as they are captured. These differentially replenished tissues will represent the dietary proportions consumed over different periods. For example, whilst blood plasma might represent the immediately sampled food sources, feathers might represent the diet consumed over the previous few months. A long-term data set where multiple tissues are analysed simultaneously may allow for increased precision in the dietary proportions. Alternatively, it may be possible to estimate the time scale over which the tissues are responding.
\item Clustering/Mixture models to determine groupings. If there are hidden groupings amongst the organisms it may be possible to discern them using a model-based clustering approach  \citep[sensu][]{Fraley2002}. Even without such groupings, increased flexibility can be obtained by using mixtures of Gaussian distributions to model non-parametric behaviour.
\item Long-tailed multivariate distributions to account for outliers or small sample sizes. Clearly where sample sizes are small the multivariate Gaussian assumption (especially for sources) may be invalid, and thus heavier-tailed distributions may be required. One such which seems to be most easily fitted is the Multivariate Normal-Inverse Gaussian \citep[MNIG][]{Barndorff-Nielsen1997} which has been used previously in clustering and financial settings.
\end{itemize}

These are just three of the active areas of research to which SIMMs are being applied by our group. Many others are likely to appear as the field advances. We hope to report soon on these exciting new developments.\\

\bibliography{/Users/aparnell/Dropbox/bibtex/library}
\bibliographystyle{chicago}

\end{document}